\documentclass[12pt]{article}
\usepackage{amsmath}
\usepackage{amssymb}
\newtheorem{theorem}{Theorem}[section]
\newtheorem{lemma}[theorem]{Lemma}

\newtheorem{corollary}[theorem]{Corollary}

\numberwithin{equation}{section}
\usepackage{latexsym}

\def\R{{\bf R}}

\def\x{{\vec{x}}}

\def\mfr#1/#2{\hbox{$\frac{{#1}}{{#2}}$}}
\def\uprho{\raise1pt\hbox{$\rho$}}
\def\upchi{\raise1pt\hbox{$\chi$}}
\def\dlambda{\lower1pt\hbox{$\lambda$}}
\newcommand{\xij}{|x_i-x_j|}
\begin{document}

\title{{\bf The Ground State Energy of a Dilute Bose Gas}\thanks{This 
is a {\it corrected version} of a contribution to the Proceedings of the International Conference on Partial Differential 
Equations and Mathematical Physics, Birmingham, Alabama, March 15--19, 1999,
ed. by R. Weikert and G. Weinstein, pp. 296--306 
(Amer. Math. Soc./International Press 2000)}}

\author{Elliott H. Lieb\thanks{Supported in part
by NSF Grant PHY-98 20650.}\\
Departments of Mathematics and Physics\\
Princeton University, Princeton, New Jersey 08544-0708\\ {\it
lieb@math.princeton.edu}
\and
Jakob Yngvason\thanks{Supported in part by
the Adalsteinn Kristj\'ansson Foundation, University of Iceland.} 
\thanks{Copyright by the authors. Reproduction of this article, in its
entirety, by any means, is permitted for non-commercial purposes.
}\\ Institut 
f\"ur Theoretische Physik,
Universit\"at 
Wien,\\Boltzmanngasse 5, A 1090 Vienna, Austria\\ {\it
yngvason@thor.thp.univie.ac.at}}

%\date{October 15, 1999}
\maketitle
%    General info
%\subjclass{Primary 81V10, 81V70; Secondary 81V45, 81T08, 81T16}

\begin{abstract} According to a formula that was put forward many decades ago
the ground state energy per particle of an interacting, dilute Bose gas at
density $\rho$ is $2\pi\hbar^2\rho a/m$ to leading order in $\rho a^3\ll 1$,
where $a$ is the scattering length of the interaction potential and $m$ the
particle mass.  This result, which is important for the theoretical 
description
of current experiments on Bose-Einstein condensation, has recently been
established rigorously for the first time.  We give here an account of the 
proof
that applies to nonnegative, spherically symmetric potentials decreasing 
faster
than $1/r^3$ at infinity.  \end{abstract}

\maketitle
%%%%% Type your manuscript here %%%%%
\section{Introduction}
Recent progress in the trapping and cooling of atoms has 
made the ground state properties of dilute, interacting Bose gases accessible 
to 
experimental study \cite{TRAP}, \cite{DGPS}. In the theoretical description of 
such experiments an old 
formula for the ground state energy  plays an important role. This formula, 
stated precisely in (\ref{basic}) below, is the subject of the present 
contribution, which is 
essentially an exposition of the paper \cite{LY1998}, 
incorporating some new results from \cite{LSY1999} and \cite{S1999}.

We consider the Hamiltonian for $N$ Bosons of mass 
$m$ enclosed in a cubic box $\Lambda$ of side length $L$ and interacting by a
spherically symmetric pair potential 
$v(|\x_i - \x_j|)$:
\begin{equation}\label{ham}
H_{N} = - \mu\sum_{i=1}^{N} \Delta_i + 
\sum_{1 \leq i < j \leq N} v(|\x_i - \x_j|).
\end{equation}
Here  $\x_i\in\mathbb R^3$, $i=1,\dots,N$ are the positions of the particles,
$\Delta_i$ the Laplacian with respect to $\x_{i}$,
and we have denoted ${\hbar^2}/{ 2m}$ by $\mu$ for short. (By choosing 
suitable units $\mu$ could, of course, be eliminated, but we want to keep 
track 
of the dependence of the energy on  Planck's constant and the mass.) The 
Hamiltonian
(\ref{ham}) operates on {\it symmetric} wave functions in $L^2(\Lambda^{N}, 
d\x_1\cdots d\x_N)$ as is appropriate for Bosons. The interaction 
potential will be assumed to be {\it nonnegative} and to decrease 
faster than $1/r^3$ at infinity.
 
We are interested in the ground state energy $E_{0}(N,L)$ of (\ref{ham}) in 
the 
{\it thermodynamic limit} when $N$ and $L$ tend to infinity with the 
density $\rho=N/L^3$ fixed. The energy per particle in this limit
\begin{equation} e_{0}(\rho)=\lim_{L\to\infty}E_{0}(\rho L^3,L)/(\rho 
L^3).\end{equation}
Our results about $e_{0}(\rho)$ are based on estimates on
$E_{0}(N,L)$
for finite $N$ and $L$, which are important, e.g., for the considerations of 
inhomogeneous systems in \cite{LSY1999}. 
To define  $E_{0}(N,L)$ precisely one 
must specify the boundary conditions. These should not matter for the
thermodynamic limit. 
To be on the safe side we use Neumann boundary conditions for the 
lower bound, and Dirichlet boundary conditions for the upper bound 
since these lead, respectively, to the lowest and the highest energies.

For experiments with dilute gases the {\it low density asymptotics} of 
$e_{0}(\rho)$ is of importance. Low density means here that the mean 
interparticle distance, $\rho^{-1/3}$ is much larger than the 
{\it scattering length} $a$ of the potential, defined as
\begin{equation}a=\lim_{r\to\infty}r-\frac{u_{0}(r)}{u_{0}'(r)},\end{equation}
where $u_{0}$ solves the zero energy scattering equation,
\begin{equation}\label{scatteq}
-2\mu u_{0}^{\prime\prime}(r)+ v(r) u_{0}(r)=0
\end{equation}
with $u_{0}(0)=0$. (The factor $2$ in (\ref{scatteq}) comes from 
the reduced mass of the two particle problem.) Our main result is a 
rigorous proof of the formula
\begin{equation} e_{0}(\rho)\approx4\pi\mu\rho a\end{equation}
for $\rho a^3\ll 1$, more precisely of
\begin{theorem}[Low density limit of the ground state energy]
\begin{equation}\label{basic}
\lim_{\rho a^3\to 0}\frac {e_{0}(\rho)}{4\pi\mu\rho a}=1.	
\end{equation}	
\end{theorem}
This formula is independent of the boundary conditions used for the 
definition of $e_{0}(\rho)$.

The genesis of an understanding of $e_{0}(\rho)$ was the pioneering 
work \cite{BO} of Bogol\-iubov, and in the 50's and early 60's several
derivations of (\ref{basic}) were presented \cite{Lee-Huang-YangEtc}, 
\cite{Lieb63}, even including higher order terms:
\begin{equation}\frac{e_{0}(\rho)}{4\pi\mu\rho a}=
1+\mfr{128}/{15\sqrt \pi}(\rho a^3)^{1/2}
+8\left(\mfr{4\pi}/{3}-\sqrt 3\right)(\rho a^3)\log (\rho a^3)
+O(\rho a^3)
\end{equation}
These early developments are reviewed in \cite{EL2}. They all rely 
on some special assumptions about the ground state that have never been 
proved, or on the selection of special terms from a perturbation series 
which likely diverges. The only rigorous estimates of this period were 
established by Dyson, who derived the following bounds in 1957 for a 
gas of hard spheres \cite{dyson}: 
\begin{equation} \frac1{10\sqrt 2} \leq
	\frac{e_{0}(\rho)}{ 4\pi\mu\rho a}\leq\frac{1+2 Y^{1/3}}{ 
(1-Y^{1/3})^2}
\end{equation}
with $Y=4\pi\rho a^3/3$. While the upper bound has the asymptotically 
correct form, the lower bound is off the mark by a factor of about 1/14.
But for about 40 years this was the best lower bound available!

Since (\ref{basic}) is a basic result about the Bose gas it is clearly 
important to derive it rigorously and in reasonable generality, in 
particular for more general cases than hard spheres.  The
question immediately arises for which interaction potentials one 
may expect it to be true. A notable fact is that it
{\it not true for all} $v$ with $a>0$, since there are two body 
potentials with positive scattering length that allow many body bound 
states \cite{BA}. Our proof, presented in the sequel,  works for nonnegative 
$v$, but we 
conjecture that (\ref{basic}) holds if $a>0$ and $v$ has no $N$-body bound 
states for any $N$. The lower bound is, of course, the hardest part, but the 
upper bound is not altogether trivial either.

%%%%%%%%
Before we start with the estimates a simple computation and some 
heuristics may be helpful to make 
(\ref{basic}) plausible and motivate the formal proofs.

With  $u_{0}$ the scattering solution and 
$f_{0}(r)=u_{0}(r)/r$,
partial integration gives
\begin{eqnarray}\label{partint}
\int_{|\x|\leq R}\{2\mu|\nabla f_{0}|^2+v|f_{0}|^2\}d\x&=&
4\pi\int_{0}^{R}\{2\mu[u_{0}'(r)-(u_{0}(r)/r)]^2+v(r)|u_{0}(r)]^2\}dr
\nonumber\\&=&
8\pi\mu a |u_{0}(R)|^2/R^2\to 8\pi\mu a\quad\mbox{\rm for $R\to\infty$},
\end{eqnarray}
if $u_{0}$ is normalized so that $f_{0}(R)\to 1$ as $R\to\infty$.
Moreover, for positive interaction potentials the scattering solution 
minimizes 
the quadratic form in (\ref{partint}) for each $R$ with 
$u_{0}(0)=0$ and $u_{0}(R)$ fixed as boundary conditions. Hence the energy 
$E_{0}(2,L)$ of two 
particles in a large box, i.e., $L\gg 
a$, is approximately $8\pi\mu a/L^3$. If the gas is sufficiently 
dilute it is not unreasonable to expect that the energy is essentially 
a sum of all such two particle contributions. Since there are 
$N(N-1)/2$ pairs, we are thus lead to $E_{0}(N,L)\approx 4\pi\mu a 
N(N-1)/L^3$, which gives (\ref{basic}) in the thermodynamic limit.

This simple heuristics is far from a rigorous proof, however, 
especially for the lower bound. In fact, it is  rather remarkable that 
the same asymptotic formula holds both for `soft' interaction 
potentials, where perturbation theory can be expected to be a good 
approximation, and potentials like hard spheres where this is not so.
In the former case the ground state is approximately the constant 
function and the energy is {\it mostly potential}:
According to perturbation theory
$E_{0}(N,L)\approx  N(N-1)/(2 L^3)\int v(|\x|)d\x$. In particular it is {\it 
independent of} $\mu$, i.e. of Planck's constant and mass. Since, 
however, $\int v(|\x|)d\x$ is the first Born approximation to $8\pi\mu 
a$ (note that $a$ depends on $\mu$!), this is not in conflict with 
(\ref{basic}).
For `hard' potentials on the other hand, the ground state is {\it 
highly correlated}, i.e., it is far from being a product of single 
particle states. The energy is here {\it mostly kinetic}, because the 
wave function is very small where the potential is large. These two 
quite different regimes, the potential energy dominated one and the 
kinetic energy dominated one, cannot be distinguished by the low 
density asymptotics of the energy. Whether they behave 
differently with respect to other phenomena, e.g., Bose-Einstein 
condensation, is not known at present.

Bogolubov's analysis \cite{BO} presupposes the existence of Bose-Einstein
condensation. Nevertheless, it is correct (for the energy) for the
one-dimensional delta-function Bose gas \cite{LL}, despite the fact
that there is (presumably) no condensation in that case. It turns
out that BE condensation is not really needed in order to understand
the energy.  As we shall see,  `global' condensation can be replaced
by a `local' condensation on boxes whose size is independent of
$L$. It is this crucial understanding that enables us to prove Theorem
1.1 without having to decide about BE condensation.

An important idea of Dyson was to transform the hard sphere 
potential into a soft potential at the cost of sacrificing the 
kinetic energy, i.e., effectively to move from one 
regime to the other. We shall make use of this idea in our proof
of the lower bound below. But first we discuss the simpler upper 
bound, which relies on other ideas from Dyson's beautiful paper \cite{dyson}.

%%%%%%%%%%%
\section{Upper bound}

The following generalization of Dyson's upper bound holds 
\cite{LSY1999}, \cite{S1999}:
\begin{theorem}[Upper bound] Define $\rho_{1}=(N-1)/L^3$ and 
$b=(4\pi\rho_{1}/3)^{-1/3}$. For nonnegative potentials $v$, and $b>a$ 
the ground state energy of (\ref{ham}) with periodic boundary conditions 
satisfies
\begin{equation}\label{upperbound}
E_{0}(N,L)/N\leq 4\pi \mu \rho_{1}a\frac{1-\frac{a} 
{b}+\left(\frac{a} 
{b}\right)^2+\frac12\left(\frac{a} 
{b}\right)^3}{\left(1-\frac{a} 
{b}\right)^8}.	
\end{equation}
For Dirichlet boundary conditions the estimate holds with ${\rm 
(const.)}/L^2$ added to the right side.
Thus in the thermodynamic limit and for all boundary conditions
\begin{equation}
\frac{e_{0}(\rho)}{4\pi\mu\rho a}\leq\frac{1-Y^{1/3}+Y^{2/3}-\mfr1/2Y}
{(1-Y^{1/3})^8}.
\end{equation}
provided $Y=4\pi\rho a^3/3<1$.
\end{theorem}
{\it Remark.} The bound (\ref{upperbound}) holds for potentials 
with infinite range, provided $b>a$. For potentials of finite range 
$R_{0}$ it can be improved for $b>R_{0}$ to
\begin{equation}\label{upperbound2}
E_{0}(N,L)/N\leq 4\pi \mu \rho_{1}a\frac{1-\left(\frac{a} 
{b}\right)^2+\frac12\left(\frac{a} 
{b}\right)^3}{\left(1-\frac{a} 
{b}\right)^4}.	
\end{equation}

{\it Proof.}
We first remark that the expectation value of (\ref{ham}) with any 
trial wave function gives an upper bound to the bosonic ground state 
energy, even if the trial function is not symmetric under permutations 
of the variables.  The reason is that an absolute ground state of the 
elliptic differential operator (\ref{ham}) (i.e. a ground state 
without symmetry requirement) is a nonnegative function which can be 
be symmetrized without changing the energy because (\ref{ham}) is 
symmetric under permutations.  In other words, the absolute ground 
state energy is the same as the bosonic ground state energy.

Following \cite{dyson} we choose a trial function of
the following form
\begin{equation}\label{wave}
\Psi(x_1,\dots,x_N)=F_1(x_{1}) \cdot F_2(x_{1},x_{2}) \cdots 
F_N(x_{1},\dots,x_{n}).
\end{equation}
More specifically, $F_{1}\equiv 1$ and $F_{i}$
depends only on the distance of $x_{i}$ to its nearest neighbor among 
the the points  $x_1,\dots ,x_{i-1}$ (taking the periodic boundary 
into account):
\begin{equation}\label{form}
F_i(x_1,\dots,x_i)=f(t_i), 
\quad t_i=\min\left(\xij,j=1,\dots, 
i-1\right),
\end{equation}
with a function $f$ satisfying  
\begin{equation}0\leq 
f\leq 1, \quad f'\geq 0.
\end{equation}
The intuition behind the ansatz (\ref{wave}) is that the particles are 
inserted into the system one at the time, taking into account the 
particles previously inserted. While such a wave function cannot 
reproduce all correlations present in the true ground state, it turns 
out to capture the leading term in the energy for dilute gases. 
The form (\ref{form}) is  computationally easier to handle than an 
ansatz of the type $\prod_{i<j}f(|x_{i}-x_{j}|)$, which might appear 
more natural in view of the heuristic remarks at the end of the last 
section.

The function $f$ is chosen to be
\begin{equation}\label{deff}
f(r)=\begin{cases}
f_0(r)/f_{0}(b)&\text{for $0\leq r\leq b$},\\
1&\textrm{for $r>b$},
\end{cases}
\end{equation}
with $f_{0}(r)=u_{0}(r)/r$. The estimates (\ref{upperbound}) and 
(\ref{upperbound2}) are 
obtained by somewhat lengthy computations similar as in 
\cite{dyson}, but making use of 
(\ref{partint}). For details we refer to \cite{LSY1999} and \cite{S1999}.

A test wave function with Dirichlet boundary condition may be obtained 
by localizing the wave function (\ref{wave}) on the length scale $L$. 
The energy cost per particle for this is ${\rm (const.)}/L^2$.
\hfill$\Box$	

\section{Lower bound}
%%%%%%%%%%%%%%%%%%%%%%
To get an idea why the lower bound
 for the bosonic ground state energy of (\ref{ham}) is not easy to 
 obtain let us consider the relevant length scales of the problem.
 These are
\begin{itemize}
\item The scattering length $a$.
\item The mean particle distance $\rho^{-1/3}$.
\item The `uncertainty principle length' $\ell_{c}$, defined by
$\mu\ell_{c}^{-2}=e_{0}(\rho)$, i.e., $\ell_{c}\sim (\rho a)^{-1/2}$.
\end{itemize}
The length $\ell_{c}$ is sometimes called `correlation length' or 
`healing length'. The name  `uncertainty principle length' 
is justified by the fact that this is the shortest length scale on 
which the bosons can be localized without raising the energy per 
particle above 
$e_{0}$, according to the uncertainty principle. For dilute gases
$\rho a^3\ll 1$ and hence
\begin{equation}
a\ll \rho^{-1/3}\ll (\rho a^3)^{-1/6}\rho^{-1/3}\sim \ell_{c}.	
\end{equation}
Bosons in their ground state are therefore `smeared out' over distances 
large compared to the mean particle distance and their individuality 
is entirely lost. Fermions, on the other hand, prefer to sit in 
private rooms, i.e., $\ell_{c}$ can be comparable to $\rho^{-1/3}$. 
In this respect the quantum nature of Bosons is much more pronounced 
than for Fermions.
The three different length scales for Bosons will play a role in the 
proof below.

Our lower bound for $e_{0}(\rho)$ is as follows.
\begin{theorem}[Lower bound in the thermodynamic limit]\label{lbth}  
For a  positive potential $v$ with finite range and $Y$ small enough
\begin{equation}\label{lowerbound}\frac{e_{0}(\rho)}{4\pi\mu\rho a}\geq 
(1-C\, 
Y^{1/17})
\end{equation}
with $C$ a constant. If $v$ does not have finite range, but decreases at 
least as fast as 
$1/r^{3+\varepsilon}$ at infinity with some $\varepsilon>0$, then an analogous 
bound to (\ref{lowerbound})
holds, but with $C$ replaced by another constant and 1/17 by another 
exponent, both of which may depend on $\varepsilon$.
\end{theorem}
It should be noted right away that the error term $-C\, Y^{1/17}$ in
(\ref{lowerbound}) is of no fundamental significance and is
not believed to reflect the true state of affairs. Presumably, it 
does not even have the right sign. We mention in passing that
$C$ can be taken to be
$8.9$ \cite{S1999}.

As mentioned in the Introduction a lower bound on $E_{0}(N,L)$ for 
finite $N$ and $L$ is of importance for applications to inhomogeneous 
gases, and in fact we derive (\ref{lowerbound}) from such a bound. We 
state it in the following way:
\begin{theorem}[Lower bound in a finite box] \label{lbthm2} 
	For a  positive potential $v$ with finite range there is 
a $\delta>0$ such that the the ground state energy of (\ref{ham}) with Neumann 
conditions satisfies
\begin{equation}\label{lowerbound2}E_{0}(N,L)/N\geq 4\pi\mu\rho 
a \left(1-C\, 
Y^{1/17}\right)
\end{equation} 
for all $N$ and $L$ with $Y<\delta$ and $L/a>C'Y^{-6/17}$. Here 
$C$ and $C'$ are constants,
independent of $N$ and $L$. (Note that the condition on $L/a$
 requires in particular that $N$ must be large enough, 
 $N>\hbox{\rm (const.)}Y^{-1/17}$.) 
 As in Theorem \ref{lbth} such a bound, but possibly with other 
 constants and another 
 exponent for $Y$, holds also for potentials $v$ of infinite range
 decreasing faster than $1/r^3$ at infinity.
\end{theorem}

The first step in the proof of (\ref{lbthm2}) is a generalization of 
a lemma of Dyson, which allows us to replace $v$ by a `soft' potential, 
at the cost of sacrificing kinetic energy and increasing the 
effective range.

\begin{lemma}\label{dysonl} Let $v(r)\geq 0$ with finite range $R_{0}$. Let 
$U(r)\geq 0$ 
be any function satisfying $\int U(r)r^2dr\leq 1$ and $U(r)=0$ for $r<R_{0}$. 
Let 
${\mathcal B}\subset \R^3$ be star shaped with respect to $0$ (e.g.\ 
convex with $0\in{\mathcal B}$). Then for all differentiable 
functions $\psi$
\begin{equation}\label{dysonlemma}
	\int_{\mathcal B}\left[\mu|\nabla\psi|^2+\mfr1/2 
v|\psi|^2\right]
\geq \mu a \int_{\mathcal B} U|\psi|^2.\end{equation}
\end{lemma}
{\it Proof.}  Actually, (\ref{dysonlemma}) holds with $\mu |\nabla \phi 
(\x)|^2$
replaced by the (smaller) radial kinetic energy,
 $\mu |\partial \phi (\x)/ \partial r|^2$, and  it  suffices to 
prove
the analog of (\ref{dysonlemma}) for the integral along each radial
line with fixed angular variables. Along such a line we write 
$\phi(\x) = u(r)/r$ with $u(0)=0$. We consider first the special case 
when when $U$ is a delta-function at some radius $R\geq 
R_0$,
i.e., \begin{equation}\label{deltaU}U(r)=\frac{1}{ 
R^2}\delta(r-R).\end{equation}
For such $U$ the analog of (\ref{dysonlemma}) along the radial line is
\begin{equation}\label{radial}\int_{0}^{R_{1}}
	\{\mu[u'(r)-(u(r)/r)]^2+\mfr1/2v(r)|u(r)]^2\}dr\geq
	\begin{cases}
		0&\text{if $R_{1}<R$}\\
			\mu a|u(R)|^2/R^2&\text{if $R\leq R_{1}$}
\end{cases}
\end{equation}
where $R_{1}$ is the length of the radial line segment in ${\mathcal 
B}$.
The case $R_{1}<R$ is trivial, 
because $\mu|\partial \psi/\partial r|^2+\mfr1/2 v|\psi|^2\geq 0$. 
(Note that positivity of $v$ is used here.) If $R\leq R_{1}$ we 
consider the integral on the the left side of (\ref{radial}) from 0 to $R$ 
instead of $R_{1}$ and
minimize it under the boundary condition that $u(0)=0$ 
and $u(R)$ is a fixed constant. Since everything is homogeneous in $u$ we may 
normalize this value to $u(R)=R-a$. 
This minimization problem leads to the zero energy 
scattering 
equation (\ref{scatteq}). Since $v$ is positive, the
solution is a true minimum and not just a 
stationary point.

Because $v(r)=0$ for $r>R_{0}$ the solution, $u_{0}$, satisfies $u_{0}(r)=r-a$ 
for $r>R_{0}$.  By partial integration, 
\begin{equation}\int_{0}^{R}\{\mu[u'_{0}(r)-(u_{0}(r)/r)]^2+
	\mfr1/2v(r)|u_{0}(r)]^2\}dr=\mu a|R-a|^2/R^2.
	\end
{equation}
But $|R-a|^2/R^2$ is precisely 
the right side of (\ref{radial}) if $u$ satisfies the normalization condition.

This derivation of (\ref{dysonlemma}) for the special case (\ref{deltaU}) 
implies the 
general case, because every $U$ can be written as a 
superposition of  $\delta$-functions, 
$U(r)=\int R^{-2}\delta(r-R)\,U(R)R^2 dR$, and $\int U(R)R^2 dR\leq 1$ 
by assumption. 
\hfill$\Box$

By dividing $\Lambda$ for given points $\x_{1},\dots,\x_{N}$ into Voronoi 
cells  ${\mathcal B}_{i}$ that contain all points 
closer to $\x_{i}$ than to $\x_{j}$ with $j\neq i$ (these 
cells are star shaped w.r.t. $\x_{i}$, indeed convex), the 
following corollary of Lemma \ref{dysonl} can be derived in the same 
way as the corresponding  Eq.\ (28) in \cite{dyson}.

\begin{corollary} For any $U$ as in Lemma \ref{dysonl}
\begin{equation}\label{corollary}H_{N}\geq \mu a W\end{equation}
with
\begin{equation}\label{W}W(\x_{1},\dots,\x_{N})=\sum_{i=1}^{N}U(t_{i}),
\end{equation}
where $t_{i}$ is the distance of $\x_{i}$ to its {\it nearest 
neighbor} among the other points $\x_{j}$, $j=1,\dots, N$, i.e.,
\begin{equation}t_{i}(\x_{1},\dots,\x_{N})=\min_{j,\,j\neq 
i}|\x_{i}-\x_{j}|.\end{equation}
\end{corollary}
\noindent
(Note that $t_{i}$ has here a slightly different meaning than in 
(\ref{form}), where it denoted the distance to the nearest neighbor 
among the $\x_{j}$ with $j\leq i-1$.)

Dyson considers in \cite{dyson} a one parameter family of $U$'s that 
is essentially the same as the following choice, which is convenient for the 
present purpose:
\begin{equation}U_{R}(r)=\begin{cases}3(R^3-R_{0}^3)^{-1}&\text{for 
$R_{0}<r<R$ }\\
0&\text{otherwise.}
\end{cases}
\end{equation} 
We denote the corresponding interaction (\ref{W}) by $W_R$. For the hard core 
gas one obtains
\begin{equation}\label{infimum}E(N,L)\geq \sup_{R}\inf_{(\x_{1},\dots,\x_{N})} 
\mu a
W_R(\x_{1},\dots,\x_{N})\end{equation}
where the infimum is over $(\x_{1},\dots,x_{N})\in\Lambda^{N}$ with 
$|\x_{i}-\x_{j}|\geq R_{0}=a$, 
because of the hard core. At fixed $R$ simple geometry gives
\begin{equation}\label{fixedR}\inf_{(\x_{1},\dots,\x_{N})}
W_R(\x_{1},\dots,\x_{N})\geq \left(\frac{A}{R^3}-\frac{B}{ \rho 
R^6}\right)\end{equation}
with certain constants $A$ and $B$. An evaluation of these constants 
gives Dyson's bound
\begin{equation}E(N,L)/N\geq \frac{1}{10\sqrt 2} 4\pi\mu \rho 
a.\end{equation}

The main reason this method does not give a better bound is that $R$ 
must be chosen quite big, namely of the order of the mean particle 
distance $\rho^{-1/3}$, in order to guarantee 
that the spheres 
of radius $R$ around the $N$ points overlap. Otherwise the infimum of 
$W_R$ will be zero. But large $R$ means that $W_R$ is 
small. It should also be noted that this method does not work 
for potentials other than hard spheres: If $|\x_{i}-\x_{j}|$ 
is allowed to be less than $R_{0}$, then the right side of 
(\ref{infimum}) is zero because $U(r)=0$ for $r<R_{0}$.

For these reasons we take another route.
We still use  Lemma \ref{dysonlemma} to get into the soft potential regime, 
but we do {\it  not} 
sacrifice {\it all} the 
kinetic energy as in (\ref{corollary}). Instead we
write, for $\varepsilon>0$
\begin{equation}
	H_{N}=\varepsilon H_{N}+(1-\varepsilon)H_{N}\geq \varepsilon 
T_{N}+(1-\varepsilon)H_{N}
\end{equation}
with $T_{N}=-\sum_{i}\Delta_{i}$ and use (\ref{corollary}) only for the
part $(1-\varepsilon)H_{N}$. This gives
\begin{equation}\label{halfway}H_{N}\geq \varepsilon T_{N}+(1-\varepsilon)\mu 
a
W_R.\end{equation}
We consider the operator on the right side 
from the viewpoint of first order perturbation theory, 
with $\varepsilon T_{N}$ as the unperturbed  part, denoted $H_{0}$.

The ground state of $H_{0}$ in a box of side length $L$ is
$\Psi_{0}(\x_{1},\dots,\x_{N})\equiv L^{-3N/2}$ and we denote 
expectation values in this state by $\langle\cdot\rangle_{0}$.
A  computation, cf.\ Eq.\ (21) in \cite{LY1998}, gives
\begin{eqnarray}\label{firstorder}4\pi\rho\left(1-\mfr1/N\right)&\geq&
\langle W_R\rangle_{0}/N\\ &\geq& 4\pi\rho
\left(1-\mfr1/N\right)\left(1-\mfr{2R}/L\right)^3
\left(1+4\pi\rho(1-\mfr1/N)(R^3-R_{0}^3)/3)\right)^{-1}.\nonumber
\end{eqnarray}
The rationale behind the various factors is as follows: $(1-\mfr1/N)$ comes 
from 
the fact that the number of pairs is $N(N-1)/2$ and not $N^2/2$, 
$(1-{2R}/L)^3$ 
takes into account the fact that the particles do not interact beyond the 
boundary of 
$\Lambda$, and the last factor measures the probability to find another 
particle 
within the interaction range of the potential $U_R$ for a given particle.

The first order result (\ref{firstorder}) looks  at first sight quite  
promising, for if we let 
$L\to \infty$, $N\to \infty$ with $\rho=N/L^3$ fixed, and 
subsequently take
$R\to\infty$, then  $\langle W_R\rangle_{0}/N$ converges to $4\pi\rho$, which 
is 
just what is desired.
But the first order result (\ref{firstorder}) is not a 
rigorous bound on $E_0(N,L)$, we need
{\it error estimates}, and these will depend on $\varepsilon$, $R$ 
and $L$.

We now recall {\it Temple's inequality} \cite{TE} for the expectations 
values of an operator $H=H_{0}+V$ in the ground state 
$\langle\cdot\rangle_{0}$ of $H_{0}$. It is a simple 
consequence of the operator inequality
\begin{equation}(H-E_{0})(H-E_{1})\geq 0\end{equation}
for the two lowest eigenvalues, $E_{0}<E_{1}$, of 
$H$ and reads
\begin{equation}\label{temple}E_{0}\geq \langle H\rangle_{0}-\frac{\langle 
H^2\rangle_{0}-\langle 
H\rangle_{0}^2}{E_{1}-\langle H\rangle_{0}}\end{equation}
provided $E_{1}-\langle H\rangle_{0}>0$.
Furthermore, if $V\geq 0$ we may use $E_{1}\geq E_{1}^{(0)}$= second lowest 
eigenvalue of $H_{0}$ and replace $E_{1}$ in (\ref{temple}) by $E_{1}^{(0)}$.

%BEGIN CHANGE
{}From (\ref{firstorder}) and (\ref{temple}) we get the estimate
\begin{equation}\label{estimate2}\frac{E_{0}(N,L)}{ N}\geq 4\pi \mu a\rho
\left(1-{\mathcal 
E}(\rho,L,R,\varepsilon)\right)\end{equation}
with
\begin{eqnarray}\label{error}1-{\mathcal 
E}(\rho,L,R,\varepsilon)&=&(1-\varepsilon)\left(1-\mfr1/{\rho 
L^3}\right)\left(1-\mfr{2R}/L\right)^3
\left(1+\mfr{4\pi}/3\rho(1-\mfr1/N)(R^3-R_{0}^3))\right)^{-1}\nonumber\\ 
&\times&\left(1-\frac{\mu a\big(\langle 
W_R^2\rangle_0-\langle W_R\rangle_0^2\big)}{\langle 
W_R\rangle_0\big(E_{1}^{(0)}-\mu a\langle W_R\rangle_0\big)}\right).
\end{eqnarray}
%END CHANGE
To evaluate this further one may use the estimates (\ref{firstorder}) and the 
bound
\begin{equation}\label{square}
\langle W_R^2\rangle_0\leq 3\frac N{R^3-R_0^3}\langle W_R\rangle_0  
\end{equation}
which follows from $U_R^2=3({R^3-R_0^3})^{-1}U_R$ together with the 
Cauchy-Schwarz 
inequality. A glance at the form of the error term reveals, however, that it 
is 
{\it not} possible here to take the thermodynamic limit $L\to\infty$ with 
$\rho$ 
fixed:
We have $E_{1}^{(0)}=\varepsilon\pi\mu/L^2$ (this is the kinetic energy of a 
{\it single} particle in the first excited state in the box), and the factor
$E_{1}^{(0)}-\mu a\langle W_R\rangle_0$ in the denominator in (\ref{error}) 
is, 
up to unimportant constants and lower order terms, $\sim (\varepsilon 
L^{-2}-a\rho^2L^3)$. Hence the denominator eventually becomes negative and 
Temple's inequality looses its validity if $L$ is large enough.

As a way out of this dilemma we divide the big box $\Lambda$ into cubic {\it
cells} of side length $\ell$ that is kept {\it fixed} as $L\to \infty$.  The
number of cells, $L^3/\ell^3$, on the other hand, increases with $L$.  The $N$
particles are distributed among these cells, and we use (\ref{error}), with 
$L$
replaced by $\ell$, $N$ by the particle number, $n$, in a cell and $\rho$ by
$n/\ell^3$, to estimate the energy in each cell with {\it Neumann} conditions 
on the boundary.  This boundary condition leads to lower energy than any other
boundary condition.  For each distribution of the particles we add the
contributions from the cells, neglecting interactions across boundaries.  
Since
$v\geq 0$ by assumption, this can only lower the energy.  Finally, we minimize
over all possible choices of the particle numbers for the various cells 
adding up to $N$.  The energy obtained in this way is a lower bound to 
$E_0(N,L)$,
because we are effectively allowing discontinuous test functions for the
quadratic form given by $H_N$.

In mathematical terms, the cell method leads to 
\begin{equation}\label{sum}
E_0(N,L)/N\geq(\rho\ell^3)^{-1}\inf \sum_{n\geq 0}c_nE_0(n,\ell)
\end{equation}
where the infimum is over all choices of coefficients $c_n\geq 0$ (relative 
number of cells containing exactly $n$ particles), satisfying the constraints
\begin{equation}\label{constraints}
\sum_{n\geq 0}c_n=1,\qquad \sum_{n\geq 0}c_n n=\rho\ell^3.
\end{equation}

The minimization problem for the distributions of the particles among the 
cells would be easy if we knew that the ground state energy $E_0(n,\ell)$ (or 
a 
good
lower bound to it) were convex in $n$.  Then we could immediately conclude 
that
it is best to have the particles as evenly distributed among the boxes as
possible, i.e., $c_n$ would be zero except for the $n$ equal to the 
integer closest to 
$\rho\ell^3$. This would give 
\begin{equation}\label{estimate3}\frac{E_{0}(N,L)}{ N}\geq 4\pi \mu a\rho
\left(1-{\mathcal E}(\rho,\ell,R,\varepsilon)\right)\end{equation} i.e.,
replacement of $L$ in (\ref{estimate2}) by $\ell$, which is independent of 
$L$.
The blow up of ${\mathcal E}$ for $L\to\infty$ would thus be avoided.

Since convexity of $E_0(n,\ell)$ is not known (except in the thermodynamic 
limit) 
we must resort to other means to show that $n=O(\rho\ell^3)$ in all 
boxes. The rescue 
comes from {\it superadditivity} of $E_{0}(n,\ell)$, i.e., the property
\begin{equation}\label{superadd}
 E_0(n+n',\ell)\geq E_0(n,\ell)+E_0(n',\ell)
\end{equation}
which follows immediately from $v\geq 0$ by dropping the interactions between 
the $n$ particles and the $n'$ particles. The bound (\ref{superadd}) implies 
in 
particular that for any $n,p\in{\mathbb N}$ with $n\geq p$
\begin{equation}\label{superadd1}
E(n,\ell)\geq [n/p]\,E(p,\ell)\geq \frac n{2p}E(p,\ell)
\end{equation}
since the largest integer $[n/p]$ smaller than $n/p$ is in any case $\geq 
n/(2p)$.

The way (\ref{superadd1}) is used is as follows:
Replacing $L$ by $\ell$, $N$ by $n$ and $\rho$ by $n/\ell^3$ in 
(\ref{estimate2})  we have for fixed $R$ and $\varepsilon$
\begin{equation}\label{estimate4}
E_{0}(n,\ell)\geq\frac{ 4\pi \mu a}{\ell^3}n(n-1)K(n,\ell)
\end{equation}
with a certain function $K(n,\ell)$ determined by (\ref{error}). We 
shall see that $K$ is monotonously decreasing in $n$, so that if 
$p\in{\mathbb N}$  and $n\leq p$ then
\begin{equation}\label{n<p}
E_{0}(n,\ell)\geq\frac{ 4\pi \mu a}{\ell^3}n(n-1)K(p,\ell).
\end{equation}
We now split the sum in (\ref{sum}) into two parts. 
For $n<p$ we use (\ref{n<p}), and for $n\geq p$ we use (\ref{superadd1}) 
together with (\ref{n<p}) for $n=p$. The task is thus to minimize
\begin{equation}\label{task}
\sum_{n<p}c_n n(n-1)+\mfr1/2\sum_{n\geq p}c_nn(p-1)
\end{equation}
subject to the constraints ({\ref{constraints}). 
Putting 
\begin{equation}
k:=\rho\ell^3 \quad\text{and}\quad t:=\sum_{n<p}c_n n\leq k
\end{equation}
we have $\sum_{n\geq p}c_n n=k-t$, and since 
$n(n-1)$ is convex in $n$, and $\sum_{n<p}c_n\leq 1$ the expression 
(\ref{task})
is
\begin{equation}
\geq t(t-1)+\mfr1/2(k-t)(p-1).
\end{equation}
We have to minimize this for $1\leq t\leq k$. If $p\geq 4k$ the minimum is 
taken 
at $t=k$ and is equal to $k(k-1)$. Altogether we have thus shown that
%BEGIN CHANGE
\begin{equation}\label{estimate1}
\frac{E_{0}(N,L)}{ N}\geq 4\pi \mu a\rho\left(1-\frac1{\rho\ell^3} \right) 
K(4\rho\ell^3,\ell).
\end{equation}
%END CHANGE

What remains is to take a closer look at $K(4\rho\ell^3,\ell)$, which depends 
on 
the parameters $\varepsilon$ and $R$ besides $\ell$, and choose the parameters 
in an optimal way. 
%BEGIN CHANGE
>From (\ref{error}) and 
(\ref{square}) we obtain
\begin{eqnarray}\label{Kformula}
K(n,\ell)&=&(1-\varepsilon) \left(1-\mfr{2R}/\ell\right)^3
\left(1+\mfr{4\pi}/3\rho(1-\mfr1/n)(R^3-R_{0}^3))\right)^{-1}
\nonumber
\\ &\times&\left(1-\frac3\pi
\frac{an}{(R^3-R_{0}^3)(\varepsilon\ell^{-2}-4a\ell^{-3}n(n-1))}\right).	
\end{eqnarray}
The estimate (\ref{estimate4}) with this $K$ is valid as long as the 
denominator in the last factor
%END CHANGE
in (\ref{Kformula}) is $\geq 0$, and in order to have a formula 
for 
all $n$ we can take 0 as a 
trivial lower bound in other cases or when (\ref{estimate4}) is 
negative. As required
for (\ref{n<p}), $K$ is monotonously decreasing in $n$. We now insert
$n=4\rho\ell^3$ and obtain
%BEGIN CHANGE
\begin{eqnarray}\label{Kformula2}
K(4\rho\ell^3,\ell)&\geq&(1-\varepsilon)\left(1-\mfr{2R}/\ell\right)^3
\left(1+({\rm const.})Y(\ell/a)^3 (R^3-R_{0}^3)/\ell^3\right)^{-1}
\nonumber
\\ &\times&\left(1-
\frac{\ell^3}{(R^3-R_{0}^3)}\frac{({\rm const.})Y}
{(\varepsilon(a/\ell)^{2}-({\rm const.})Y^2(\ell/a)^3)}\right)	
\end{eqnarray}
with $Y=4\pi\rho a^3/3$ as before. Also, the factor 
\begin{equation}
\left(1-\frac1{\rho\ell^3} \right)=(1-({\rm const.})Y^{-1}(a/\ell)^{3})
\end{equation}
%END CHANGE
in (\ref{estimate1})
(which is the ratio between
$n(n-1)$ and $n^2$) must not be be forgotten. We now make the ansatz
\begin{equation}\label{ans}
\varepsilon\sim Y^\alpha,\quad a/\ell\sim Y^{\beta},\quad 
(R^3-R_{0}^3)/\ell^3\sim Y^{\gamma}	
\end{equation}	
with exponents $\alpha$, $\beta$ and $\gamma$ that we choose 
in an optimal way. The conditions to be met are as follows:
%BEGIN CHANGE
\begin{itemize}
\item $\varepsilon(a/\ell)^{2}-({\rm const.})Y^2(\ell/a)^3>0$. This 
holds for all small enough $Y$, provided 
$\alpha+5\beta<2$ which follows from the conditions below.
\item $\alpha>0$ in order that $\varepsilon\to 0$ for $Y\to 0$.
\item $3\beta-1>0$ in order that  $Y^{-1}(a/\ell)^{3}\to 0$ for for $Y\to 
0$. 
\item $1-3\beta+\gamma>0$ in order that  
$Y(\ell/a)^{3}(R^3-R_{0}^3)/\ell^3\to 0$ for for $Y\to 0$.
%END CHANGE
\item $1-\alpha-2\beta-\gamma>0$ to control the last factor in 
(\ref{Kformula2}).
\end{itemize}
Taking
\begin{equation}\label{exponents}
\alpha=1/17,\quad \beta=6/17,\quad \gamma=3/17	
\end{equation}
all these conditions are satisfied, and
%BEGIN CHANGE
\begin{equation}
\alpha=	3\beta-1=1-3\beta+\gamma=1-\alpha-2\beta-\gamma=1/17.
\end{equation}
It is also clear that 
$2R/\ell\sim Y^{\gamma/3}=Y^{1/17}$, up to higher order terms.
%END CHANGE
This completes the proof of Theorems 3.1 and 3.2, for the case 
of potentials with  finite range. By optimizing the proportionality 
constants in (\ref{ans}) one can show that $C=8.9$ is possible in Theorem 1.1 
\cite{S1999}. The extension to potentials of infinite range 
decreasing faster than $1/r^3$ at infinity is 
obtained by approximation by finite range potentials, controlling the 
change of the scattering length as the cut-off is removed. See 
Appendix B in \cite{LSY1999} for details. A slower decrease than 
$1/r^3$ implies infinite scattering length. \hfill$\Box$

The exponents (\ref{exponents}) mean in particular that
\begin{equation}a\ll R\ll \rho^{-1/3}\ll \ell \ll(\rho 
a)^{-1/2},\end{equation}
whereas Dyson's method required $R\sim \rho^{-1/3}$ as already explained. 
The condition $\rho^{-1/3}\ll \ell$ is required in order to have many 
particles in each box and thus $n(n-1)\approx n^2$. The condition
$\ell \ll(\rho a)^{-1/2}$ is necessary for a spectral gap
gap $\gg e_{0}(\rho)$ in Temple's inequality. It is also clear that 
this choice of $\ell$  would lead to a far too big
energy and no bound for $ e_{0}(\rho)$ if we had chosen Dirichlet instead of 
Neumann boundary 
conditions for the cells. But with the latter the method works!
%%%%%%%%%%


\begin{thebibliography}{99}
	
\bibitem{TRAP} W. Ketterle, N. J. van Druten, in B. Bederson, H.
Walther, eds.,  Advances in Atomic, Molecular and Optical Physics, {\bf
37}, 181, Academic Press (1996).

\bibitem{DGPS}
F. Dalfovo, S.\ Giorgini, L.P.\ Pitaevskii, and S.\ Stringari, {\it Theory of 
Bose-Einstein condensation in trapped gases},
Rev. Mod. Phys. \textbf{71}, 
463--512 (1999).

	
	
\bibitem{LY1998}
E.H. Lieb, J. Yngvason, {\it Ground State Energy of the low density Bose 
Gas}, Phys. Rev. Lett. \textbf{80}, 2504--2507 (1998).

\bibitem{LSY1999}
E.H. Lieb, R. Seiringer, and J. Yngvason, {\it Bosons in a Trap: A Rigorous 
Derivation of the Gross-Pitaevskii Energy Functional},
Phys. Rev. A {\bf 61}, 043602-1 -- 043602-13 (2000) 
%mp\_arc 99-312, xxx e-print archive math-ph/9908027 (1999).

\bibitem{S1999}
R. Seiringer, Diplom thesis, University of Vienna, 1999.



\bibitem{BO} N.N. Bogoliubov,  J.  Phys. (U.S.S.R.)  {\bf 11}, 23
(1947); N.N. Bogoliubov and D.N.  Zubarev,  Sov. Phys.-JETP {\bf 1}, 83
(1955).


\bibitem{Lee-Huang-YangEtc}K.~Huang, and C.N.~Yang, Phys. Rev. {\bf 
105}, 767-775
(1957); T.D.~Lee, K.~Huang, and C.N.~Yang,  Phys. Rev. {\bf 106}, 
1135-1145 (1957); K.A. Brueckner, K. Sawada, Phys. Rev. {\bf 106},
1117-1127, 1128-1135 (1957).; S.T. Beliaev, Sov. Phys.-JETP {\bf 7}, 
299-307 (1958); T.T. Wu, Phys. Rev. {\bf 115}, 1390 (1959); 
N. Hugenholtz, D. Pines, Phys. Rev. {\bf 116}, 489 (1959); M. 
Girardeau, R. Arnowitt, Phys. Rev. {\bf 113}, 
755 (1959); T.D. Lee,  C.N. Yang, Phys. Rev. {\bf 117}, 12 (1960).
%%
\bibitem{Lieb63} E.H. Lieb, {\it 
Simplified Approach to the Ground State Energy of an Imperfect Bose Gas}, 
Phys. Rev. {\bf 130}, 2518--2528 (1963). See also Phys. Rev. {\bf 133},
A899-A906 (1964) (with A.Y. Sakakura) and Phys. Rev. {\bf 134},
A312-A315 (1964) (with W. Liniger).

\bibitem{EL2} E.H.~Lieb, {\it The Bose fluid,} in Lecture Notes in
Theoretical Physics VIIC, W.E.~Brittin, ed., Univ. of Colorado Press,
pp. 175 (1964).

\bibitem{dyson}
F.J. Dyson, {\it Ground-State Energy of a Hard-Sphere Gas}, 
Phys. Rev. \textbf{106}, 20--24 (1957).


\bibitem{BA} B. Baumgartner,{\it The existence of many-particle bound 
states despite a pair interaction with positive scattering length}, 
J. Phys. A {\bf 30}, L741--L747 (1997).

\bibitem{LL} E.H. Lieb, W. Liniger, {\it Exact Analysis of an
Interacting Bose Gas. I.  The General Solution and the Ground State},
Phys. Rev. \textbf{130}, 1605--1616 (1963).


\bibitem{TE} G. Temple, {\it The theory of Rayleigh's principle as
applied to continuous systems},
Proc. Roy. Soc. London A \textbf{119}, 276-293 (1928).

\end{thebibliography}
\end{document}